\documentclass[12pt, letterpaper, onehalfspacing]{article}

\usepackage{libertine}
\usepackage[pdftex]{graphicx}
\usepackage{enumerate}
\usepackage[margin=1.25in]{geometry}
\usepackage{dsfont}
\usepackage{setspace}

\usepackage{booktabs} % for much better looking tables
\usepackage{array} % for better arrays (eg matrices) in maths
\usepackage{paralist} % very flexible & customisable lists (eg. enumerate/itemize, etc.)
\usepackage{verbatim} % adds environment for commenting out blocks of text & for better verbatim
\usepackage{subfig} % make it possible to include more than one captioned figure/table in a single float
\usepackage{amsfonts, amsmath, amssymb, amsthm, euscript}
\usepackage{mathrsfs}
\usepackage{mhsetup, mathtools}
\usepackage{thmtools}
\usepackage{epigraph}
\usepackage{color}
\usepackage{cancel}
\usepackage{adjustbox}
\usepackage{xcolor, graphicx} % support the \includegraphics command and options
\usepackage{tikz}
\usepackage{multicol}
\usepackage{tikz-cd}

\usetikzlibrary{patterns}

\definecolor{purple}{RGB}{85, 6,139}
\definecolor{teal}{RGB}{2,108,128}
\definecolor{lavender}{RGB}{129, 102, 122}
\definecolor{carolina blue}{RGB}{68, 157, 209}
\definecolor{phthalo blue}{RGB}{2, 8, 135}
\definecolor{purple2}{RGB}{149, 96, 219}
\definecolor{green1}{RGB}{96, 219, 117}
\definecolor{orange1}{RGB}{208,70,0}

\usepackage{natbib}
\usepackage[colorlinks]{hyperref}
\usepackage{nameref}
\hypersetup{
  colorlinks   = true, %Color links instead of ugly boxes
  urlcolor     = purple, %Colour for external hyperlinks
  linkcolor    = teal, %Colour of internal links
  citecolor   = gray%gray%red!40!black} %Colour of citations
}
\newtheorem{theo}{Theorem}

\newtheorem{defin}{Definition}
\newtheorem{prop}{Proposition}

\newtheorem{exm}{Example}

\newtheorem{ax}{Axiom}

\theoremstyle{definition}
\newtheorem{rem*}{Remark}

\DeclareMathOperator*{\argmax}{arg\,max}

\newcommand{\R}{\mathbb{R}}

\newcommand{\F}{\mathcal{F}}

\begin{document}

\title{\textsc{Ordered Surprises and Conditional Probability Systems}\thanks{Dominiak: Virginia Tech (dominiak@vt.edu); Kovach: Virginia Tech (mkovach@vt.edu); Tserenjigmid: UC Santa Cruz (gtserenj@ucsc.edu ). We are very grateful to Shachar Kariv, Burkhard Schipper, and Marie-Louise Vier\o \ for many stimulating discussions and suggestions that improved the exposition of the paper. All mistakes are our own.}}

\author{\centering Adam Dominiak  \and Matthew Kovach \and Gerelt Tserenjigmid}

\date{First version: April 4, 2021;  This version: \today}

\maketitle

\begin{abstract}
We study conditioning on null events, or \emph{surprises}, and behaviorally characterize the Ordered Surprises (OS) representation of beliefs. For feasible events, our Decision Maker (DM) is Bayesian. For null events, our DM considers a hierarchy of beliefs until one is consistent with the surprise. The DM adopts this prior and applies Bayes' rule. Unlike Bayesian updating, OS is a complete updating rule: conditional beliefs are well-defined for any event.  OS is (behaviorally) equivalent to the Conditional Probability System \citep{Myerson1986a} and is a special case of Hypothesis Testing \citep{Ortoleva2012}, clarifying the relationships between the various approaches to null events.

\vspace{7 mm}

\vspace{5 mm}
\noindent\textbf{JEL-Classifications:} D01, D80, D81, D83. 
\end{abstract}

\vspace{20 mm}

\newpage

\section{Introduction}

Decision problems under uncertainty often feature a dynamic structure. A decision maker (DM) may acquire information about the uncertainty she faces. Given this information, the DM must formulate a conditional belief that governs her conditional behavior. Bayesian subjective expected utility is the dominant theory in economics for dynamic choice problems. However, the Bayesian paradigm is incomplete; it is silent about updating on (Savage) null events (i.e., zero probability events). Our goal is to introduce a theory of conditional beliefs that are well defined for all events, thereby establishing a complete theory of updating.

We start by defining an updating rule, which is a mapping from events to conditional beliefs, and introduce two simple properties. An updating rule is \emph{complete} if it is has full domain (it is defined for every event), and it is \emph{concentrated} if, for each event, no states outside the event have positive probability. Complete updating rules are important, for example, because any perfect Bayesian equilibrium, including its refinements, requires one to describe equilibrium beliefs. 

The most prominent complete updating rule is Myerson's Conditional Probability System (CPS)\footnote{The idea of CPS goes back to \cite{Renyi1955}.} \citep{Myerson1986b, Myerson1986a}, which was motivated by the sequential equilibrium of \cite{KrepsWilson1982} (henceforth, KW).\footnote{Sequential equilibrium is one of the most widely applied solution concepts in dynamic games. For example, it has been used to study bargaining \citep{Rubinstein_1985}, cheap talk \citep{Crawford_Sobel_1982}, job search \citep{Spence1973}, advertising \citep{MiglromRoberts1986}, dividends \citep{John1985}, product quality \citep{Miller1985}, warranties \citep{Gal1989},  limit pricing \citep{MiglromRoberts1982},  social norms \citep{Bernheim1994}, lobbying \citep{Lohmann1995}, and many other topics.}  KW defined a sequential equilibrium as an assessment $(\sigma,\mu)$ (a strategy profile $\sigma$ and a system of beliefs $\mu$) that is ``consistent'' and ``sequentially rational.'' Consistency means that there is a sequence of full support beliefs converging to $\sigma$ such that the corresponding sequence of belief systems (Bayesian updates) converges to $\mu$. Sequential rationality requires that at each information set the actions prescribed by $\sigma$ maximize the conditional expected payoff given $(\sigma,\mu)$. Importantly, these conditions ensure that players' beliefs about play are \emph{well-defined at every information set}, even for information sets off the equilibrium path (e.g., null events); each player has a complete theory of belief updating. \cite{Myerson1986a} shows that consistency of beliefs in sequential equilibria is equivalent to CPS.\footnote{See also \cite{McLennan_1989a} and \citet{Battigalli_1996} for a connection between CPS and sequential equilibria.}

As an interpretation for how beliefs evolve in a sequential equilibrium, KW suggested that upon reaching an information set that is incompatible with a player's beliefs about play in the game, the player should move to their `` second most likely hypothesis'' and, if that fails, then they should implement their ``third most likely hypothesis,'' and so on. KW show that a sufficient condition for an assessment to satisfy this property is the existence of a convergent sequence of probability measures that generate the assessment.

We revisit the ``ordered hierarchy of hypotheses'' of KW and introduce the notion of \textbf{Ordered Surprises} (OS).  OS may be viewed as the minimal extension of Bayesian theory that allows for well-defined beliefs after null events.  An OS representation of a complete updating rule is given by a collection of beliefs $\mathcal{O}=\langle\mu_0, \ldots, \mu_K\rangle$, where $\mu_0$ is the initial belief and each $k>0$ is used to form some conditional belief after some null event. Each index indicates a distribution's order in the hierarchy. 

Our main result shows that OS is equivalent to CPS (\autoref{OS-CPS}); an updating rule has an OS representation if and only if it is a CPS. While one direction of this equivalence was suggested by KW, to the best of our knowledge, we are the first to formally prove this equivalence (CPS must have an ordered structure). Additionally, the explicit construction of the sequential beliefs for any CPS through the OS representation is useful in practice for those who want to find all sequential equilibria in extensive-form games. 

We provide simple behavioral foundations for OS, thereby establishing a simpler characterization for CPS. Our characterization (\autoref{OS-Foundation}) relies upon a novel axiom,  \nameref{CC},  that implies Dynamic Consistency among the ex-ante feasible events and extends this consistency to the entire hierarchy of beliefs; it imposes Dynamic Consistency within each $\mu_k$.

To illustrate OS, imagine a coin flip. The states $h$ and $t$ are the usual outcomes of heads or tails, $e$ and $e'$ denote edges where $e'$ has been warn thin, while $l_1$ and $l_2$ denote landing on a marked location, which yields the state space $S=\{h, t, e, e', l_1, l_2\}$.  Initially, the DM has belief $\mu_0(h)=\mu_0(t)=\frac12$, and treats the other states as null. 

Suppose the DM is informed that, astonishingly, the coin did not land on a face; $A=\{e, e', l_1, l_2 \}$ was realized.  While our DM now has sufficient evidence to rule out her initial beliefs, it is not clear which belief she should adopt. Suppose our DM believes that the coin landing on either of the marked locations is \emph{more impossible} than its landing on an edge. Accordingly, her conditional beliefs are $\mu_A(\{e, e'\})=1$ and $\mu_A(\{l_1, l_2\})=0$. Since the states in $\{e, e'\}$ are no longer null, they constitute a \emph{first-order surprise}. In contrast, $\{l_1, l_2\}$, remain null and constitute a higher-order surprise, where  ``order'' designates the point at which a state is considered possible. Our notion of OS extends this idea to up to $K$ orders. 

We compare OS to other approaches to conditioning on null events. In \autoref{OS=HTE}, we prove directly that OS preferences are a special case of the Hypothesis Testing Model of \citet{Ortoleva2012}, and thus so is CPS. In particular, for any OS given by $\langle\mu_0, \ldots, \mu_K\rangle$, we construct a second-order prior over these $K+1$ beliefs that ensures selection in the resulting HT is consistent with the OS structure. This argument cannot be reversed, and so HT is strictly more general than OS. 

CPS and OS representations complete Bayes' rule by disciplining belief updating on zero-probability events. Motivated by empirical evidence on belief updating (e.g., see \citet{benjamin2019}), we propose a one-parameter non-Bayesian extension of the OS representations based on the idea of HT. We show that this non-Bayesian extension of OS representations is still a special case of HT, confirming that it is a reasonable non-Bayesian extension of CPS. This extension might be useful for defining a non-Bayesian extension of sequential equilibrium.

\subsection{Related Literature}

There have been various attempts to deal with conditioning on zero-probability events. In economics, the three main approaches are (i) the conditional probability system (CPS) of \cite{Myerson1986b, Myerson1986a}, (ii) the (conditional) lexicographic probability systems (C)LPS of \citet{BBD1991}, and (iii) the hypothesis testing model (HT) of \citet{Ortoleva2012}. These are carefully discussed in \autoref{sec:cps}, \autoref{sec:lps}, and \autoref{sec:ht}, respectively. 

%Before delving into each of these approaches in detail, we first provide on overview of how each relates to the other. The main result of this paper shows the equivalence between the CPS and OS representations. Since we show that OS is a special case of HT, then every CPS is also an HT. This cannot be reversed. 

It is worth remarking that the (conditional) LPS is fundamentally distinct from CPS and HT (hence also OS). First, the CLPS rules out null events.%\footnote{On the other hand, an LPS that allows for null events may not produce well-defined conditionals for certain events, so it is not complete.} 
Second, (C)LPS behavior is incompatible with SEU, while the other three are SEU models.

There is an extensive literature that applies CPSs in dynamic games with incomplete information.  \cite{Battigalli_Siniscalchi_1999} use CPSs to describe an (epistemic) type of a player (i.e., a complete and explicit description of players' hierarchies of conditional beliefs) and show that a universal type space for CPSs always exists. In this framework, \citet{Battigalli_Siniscalchi_2002} provide an epistemic characterization of extensive-form rationalizability and backward induction. Recently, \cite{galperti2019persuasion} studies Bayesian persuasion using a CPS (OS with two beliefs).\footnote{ \citet{Tsakas_2018} derives the famous Agreement Theorem under a CPS.}

Finally, there is an abundant literature on (non-Bayesian) updating. Recently, \cite*{DKT2022a} show that HT admits a ``minimum-distance'' representation; the conditional probability $\mu_E$ that the DM selects is the element of $\Delta(E)$ that is closest to her prior $\mu$. Hence, CPS also admits a minimum-distance representation.

\section{Model}
\subsection{Setup}

We study dynamic choice in the formal framework of \citet{AA63}.  Uncertainty is described by a nonempty and finite set of states $S=\{s,\ldots,s\}$. We denote by $\Delta(S)$ the set of all probability distributions on $S$.  Let $X$ be a nonempty set of outcomes and $\Delta(X) := \Big\{ p : X \rightarrow [0,1] \mid \sum_{x \in supp(p)}p(x)=1~\text{with}~|supp(p)| \in \mathbb{N} \Big\}$ be the set of all (simple) lotteries over $X$, where $supp(p)$ denotes the support of $p$. 

Objects of choice are (Anscombe-Aumann) acts $f: S \to \Delta(X)$ that map states to lotteries. Let $\F$ denote the set of all such acts. A constant act is an act that assigns the same lottery to all states: $f(s)=p$ for all $s \in S$. Using a standard abuse of notation, we denote by $p \in \F$ the corresponding constant act. Thus, we can identify the set of lotteries with the constant acts. A subset $E\subseteq S$ is called an event. Denote by $\Sigma$ the algebra  of events generated by $S$. For an event $ E$ and acts $f, g$ we denote by $f_Eg$ a composite act that returns $f(s)$ for $s \in E$ and $g(s)$ otherwise.

A preference relation over $\mathcal{F}$, denoted by $\succsim$, describes a DM's behavior. As usual, $\succ$ and $\sim$ are the asymmetric and symmetric parts of $\succsim$, respectively. For each event $E\in \Sigma$,  $\succsim_E$ denotes the conditional preference over $\mathcal{F}$ given $E$. It governs the DM's choice upon learning $E$. The initial preference relation (before information is revealed) is denoted by $\succsim_S$. The DM's behavior is described by a family of conditional preferences $\{\succsim_{E}\}_{E\in \Sigma}$; one for each $E \in \Sigma$.

Given a belief $\mu \in \Delta(S)$ and an event $E \in \Sigma$ such that $\mu(E)>0$, we let $BU(\mu, E)(A)$ denote the Bayesian update of $\mu$ given $E$, where $BU(\mu, E)(A):= \frac{\mu(E \cap A)}{\mu(E)}$.

\begin{defin}\label{updating} An \textbf{updating rule} is a pair $(P,\mathscr{D})$ where $\mathscr{D} \subseteq \Sigma$ and $P$ is a mapping $P:\mathscr{D}  \to \Delta(S)$. 
\begin{itemize}
\item[(i)] An updating rule is \textbf{complete} if it has full domain: $\mathscr{D}=\Sigma$. 
\item[(ii)] An updating rule is \textbf{concentrated} if for any $E \in \mathscr{D}$, $P(E|E)=1$.
\end{itemize}

\end{defin}

In other words, $P$ is complete if the conditional probability $P(\cdot|E)$ is well-defined for each $E\in\Sigma$.

\begin{exm} Since $BU$ is only well-defined for $\mathscr{D}=\{E\in\Sigma \mid \mu(E)>0\}\subset \Sigma$ where $\mu$ is the prior, Bayesian updating is incomplete.
\end{exm}

Additionally, an LPS that admits null events (e.g., the LPS is not full support), results in an incomplete updating rule.\footnote{Technically, beliefs for a conditional LPS require an expanded notion of an updating rule. See \autoref{sec:lps} for further discussion.}

\begin{exm} Let $\mu \in \Delta(S)$, $\delta \in (0,1]$ and let $P^C$ denote ``Conservative BU,'' where  \[P^C(\cdot|E) = \begin{cases} \delta \mu + (1-\delta)BU( \mu, E) & \text{if } \mu(E)>0 \\
\delta \mu + (1-\delta)\frac{1}{|E|} & \text{if } \mu(E)=0. \end{cases}\]
\end{exm}

The updating rule $P^C$, a one-parameter extension of CPS, is complete but not concentrated (the resulting preferences violate consequentialism).\footnote{Similar updating rules have been studied in \cite{epstein2006} and \cite{kovach2020}.} When $\delta=1$, we have ``stubborn'' agents who never update their beliefs (similar behavior appears in \cite{Acemoglu_Como_Fagnani_Ozdaglar:2013}).

\subsection{Ordered Surprises}

In this section, we introduce OS, which captures the DM's subjective levels of relative (im)possibility. Intuitively, an event is null if the DM assigns it zero probability.  In behavioral terms, an event $E$ is \textit{null} if a DM is indifferent between any two acts that only differ on $E$.

\begin{defin} For any $A,E \in \Sigma$, an event $A$ is $\succsim_E$-null if for all $f, g\in \F$,
\begin{equation}
f_A g \sim_{E} g.
\end{equation}
Otherwise, $A$ is $\succsim_{E}$-feasible. If $A$ is $\succsim_{S}$-null, then it is (ex-ante) null.
\end{defin}

We can distinguish between types of null events by comparing the exa-ante preference $\succsim_{S}$ with conditional preferences $\succsim_{E}$ for various events. For example, consider the initial prior $\mu_0$ over $S$. We can partition $S$ into its feasible states $F_0$ and null states $S\setminus F_0$. When it is revealed that the true state is in fact an element of $S\setminus F_0$ --- the DM's (ex-ante) feasible states have been ruled out --- she formulates new beliefs: $\mu_{S\setminus F_0}$. 

Some states in $S\setminus F_0$ will have positive probability, while others remain null because they are \emph{more impossible}. The states that are now \emph{conditionally non-null} are ``first-order surprises,'' while the states that remain null are ``higher-order surprises.'' We can proceed in this fashion, sequentially revealing each surprise order, until all states have been considered. Following this idea, we formally define our notion of Ordered Surprises for updating rules. 

\begin{defin}\label{OS} A complete updating rule $P$ has an \textbf{Ordered Surprises} (OS) representation if there are probability distributions $\mu_0, \ldots, \mu_K \in \Delta(S)$ such that 
\[P(\cdot|E)=\text{BU}(\mu_{k^*}, E)\text{ where }k^*=\min\{k\le K \mid \mu_k(E)>0\},\] 
for every $E\in\Sigma$. For simplicity, we may refer to an OS representation as $\mathcal{O}=\langle \mu_0, \ldots, \mu_K \rangle$.
\end{defin}

It is without loss of generality to assume that $\mu_0, \ldots, \mu_K$ have disjoint supports. The indices designate the order in which each prior is considered. The DM begins with her prior over $S$, designated by $\mu_0$.  She is Bayesian whenever possible, and abandons $\mu_0$ only when necessary. After she observes a null event, she considers each $\mu_i$ in order, stops at the first belief that is consistent with the information, and applies Bayes' rule. The following example illustrates this process. 

\begin{exm}[Coin Flip]\label{coin}
Consider the coin flip example from the introduction. The states are $S=\{h, t, e, e', l_1, l_2\}$, where $h$ and $t$ correspond to heads or tails, $e$ and $e'$ correspond to the coin landing on an edge, where one edge is thinner than the other, while $l_1$ and $l_2$ correspond to the coin landing on precisely marked locations. These possibilities are described by the probability distributions 
\[ \mu_0(s)=\begin{cases} \frac12 & s \in \{h, t\} \\ 0 & \text{otherwise}\end{cases}; \mu_1(s)=\begin{cases} \frac78 & s = e \\ \frac18 & s =e' \\ 0 & \text{otherwise}\end{cases}; \mu_2(s)=\begin{cases} \frac12 & s \in \{l_1, l_2\} \\ 0 & \text{otherwise}\end{cases}. \]

Our DM has initial prior $\mu_0$ (e.g., $\succsim_S$ has an SEU representation $(u, \mu_0)$). Suppose she observes $A=\{e', l_1, l_2\}$. Since $\mu_0(A)=0$, Bayesian updating is not defined.  When our DM admits an Ordered Surprises representation generated by $\mathcal{O}=\langle\mu_0, \mu_1, \mu_2\rangle$, she selects a new belief that is consistent with $A$. In this case, \[P^{OS}(\cdot|A)=BU(\mu_1, A).\]

Note that $A$ intersects two ``surprise orders'': $\{e, e'\}\cap A \neq \varnothing \neq \{l_1, l_2\}\cap A$. Under the OS model, the DM selects $\mu_1$ because it is of lower order than $\mu_2$ and therefore takes precedence.

\end{exm}

\subsection{Conditional Probability System}\label{sec:cps}

Perhaps the most well-known method for handling choice conditional on (ex-ante) null-events is the conditional probability system introduced by \cite{Myerson1986a}. In this section, we show that the OS representation is equivalent to CPS.

\cite{Myerson1986a} initially defined a conditional probability system as a collection of conditional probabilities $\left(p(\cdot | E)\right)_{E\in \Sigma}$, each one on $S$, that jointly satisfy certain properties. We provide an equivalent definition using our semantic of complete updating rules.

\begin{defin}\label{cps}
A complete updating rule $P$ is a \textit{conditional probability system} (CPS) if it is concentrated and for all  $G \subseteq F \subseteq E$ where $F\neq \emptyset$, it satisfies
\begin{equation}\label{eq:cps}P(G|E)= P(G|F) \times P(F|E).\end{equation}
When $P(F| E)\neq 0$, \eqref{eq:cps} is equivalent to Baye's rule. 
\end{defin}

\begin{theo}\label{OS-CPS} A complete updating rule $P$ is a \textbf{Conditional Probability System} if and only if it has \textbf{Ordered Surprise} representation. 
\end{theo}

The construction of the OS representation from a CPS relies on the identification of a chain of nested ``surprise orders.'' The initial prior is $\mu_0=P(\cdot|S)$. We then separate $S$ into feasible and null states (if any). Let $S_1\subset S$ be the set of null states for $P(\cdot|S)$, then $\mu_1=P(\cdot|S_1)$. We continue in this way until we exhaust the state space, yielding $\mathcal{O}=\langle \mu_0, \ldots, \mu_K \rangle$. 

Since the OS and CPS updating rules are equivalent, families of OS preferences and CPS preferences are behaviorally equivalent as well because both admit an SEU representation (see \autoref{sec:osp}). 

\subsection{Hypothesis Testing Model}\label{sec:ht}

A recent and elegant addition to the literature on updating for null events is the Hypothesis Testing model of \cite{Ortoleva2012}.  In this model, a DM is described by a second-order prior $\rho$ and a plausibility threshold $\epsilon$.  Before any information is revealed, the DM selects her initial prior,  which is the most likely prior according to $\rho$. The DM is SEU with respect to the initial prior and applies Bayes' rule whenever the conditioning event is ``expected:" its probability exceeds $\epsilon$. When the event is ``unexpected,'' the DM revisits her second-order beliefs and selects a (possibly) new belief according to a maximum likelihood rule. Bayes' rule is then applied to the selected belief. 

Importantly, the notion of ``unexpected" events is endogenously determined via $\epsilon \in [0,1)$. When event $E$ is realized, it is unexpected if $\mu(E) \leq \epsilon$. When $\epsilon > 0$, the DM may exhibit non-Bayesian reactions to unexpected events. When $\epsilon=0$, only null-events are surprising and Bayes' rule is applied whenever possible.

\begin{defin} A complete updating rule $P$ has a \textbf{Hypothesis Testing} representation if there are probability distributions $\mu_0, \ldots, \mu_K \in \Delta(S)$ such that $S = \cup_{i=0}^{K}supp(\mu_i)$, a second-order prior $\rho \in \Delta(\{\mu_0, \ldots, \mu_K\})$ satisfying $\rho(\mu_0) > \rho(\mu_k)$ for all $k>0$, and an $\epsilon \in [0,1)$ such that 

\[P(\cdot|E)=\begin{cases}\text{BU}(\mu_0, E) & \text{ if }\mu_0(E) > \epsilon \\
\text{BU}(\mu_{k^*}, E) &  \text{ if }\mu_0(E) \leq \epsilon,
\end{cases}\]
where $\mu_{k^*}(E)\rho(\mu_{k^*}) > \mu_j(E)\rho(\mu_j)\text{ for all } j \neq k^*$.
\end{defin}

In the HT with $\epsilon=0$, the DM uses Bayes' rule for all ex-ante feasible events.  Since the DM only selects a new prior after null-events, this is intuitively similar to the OS model. We show in the following theorem that every OS representation of beliefs admits an HT representation with $\epsilon =0$. Consequently, OS preferences are a special case of HT preferences with $\epsilon=0$.\footnote{\cite{DominiakLee2022} develop a solution concept for signaling games using the HT model with $\epsilon=0$.}

\begin{theo}\label{OS=HTE}
 Any complete updating rule $P$ with an OS representation has an HT representation with $\epsilon=0$. \end{theo}

We prove this by taking an OS representation  $\mathcal{O}=\langle \mu_0, \ldots, \mu_K \rangle$ and constructing a second-order belief $\rho$ such that the resulting HT representation produces the same conditional beliefs. The key insight is that the relative probability between surprise orders must be sufficiently large so that the correct $\mu_k$ is selected. That is, if $j < l$, then $\frac{\rho(\mu_j)}{\rho(\mu_l)}$ must be large enough so that $\mu_l$ is only selected when $\mu_j$ is impossible. Viewed another way, there is an event-independent linear order that governs the selection from $\mathcal{O}$. 

This implication cannot be reversed as the following example illustrates. Of course, this is not surprising for HT with $\epsilon > 0$.  However, even when $\epsilon =0$, HT preferences may be inconsistent with OS preferences. This is because in HT, selection of a new prior $\mu$ after an event $E$ is jointly determined by ``how likely the event is under the prior,'' given by $\mu(E)$, and ``how likely the DM finds the prior,'' given by $\rho(\mu)$.

\begin{exm}\label{ex: HT} Recall our coin example (\autoref{coin}), but suppose instead that $P$ has a HT representation with $\epsilon=0$, $\rho(\mu_0)=\frac{1}{2}$, $\rho(\mu_1)=\frac{1}{3}$, and $\rho(\mu_2)=\frac{1}{6}$. When $A=\{e', l_1, l_2\}$, 
\[\mu_1(A)\rho(\mu_1)=\frac{1}{8}\cdot \frac{1}{3}< \frac{1}{6}= \mu_2(A)\rho(\mu_2),\]
since $\mu_1(A)=\frac18$ and $\mu_2(A)=1$. Thus, $P^{HT}(\cdot| A)=BU(\mu_2, A)$. However, when $E=\{e, e', l_1, l_2\}$,   
\[\mu_1(E)\rho(\mu_1)=\frac{1}{3}>\frac{1}{6} = \mu_2(E)\rho(\mu_2)\]
since $\mu_1(E)=\mu_2(E)=1$. Thus, $P^{HT}(\cdot| E)=BU(\mu_1, E)$. 
Hence, the described HT representation does not admit an OS representation because 
\[P^{HT}(e' \mid E) =\frac{1}{8} \neq 0 \cdot \frac{1}{8}= P^{HT}(e'\mid A)  \cdot P^{HT}(A\mid E),
\]
violates \eqref{eq:cps} in \autoref{cps}.
\end{exm}

\begin{figure}[h]
\centering
\includegraphics[width=6cm]{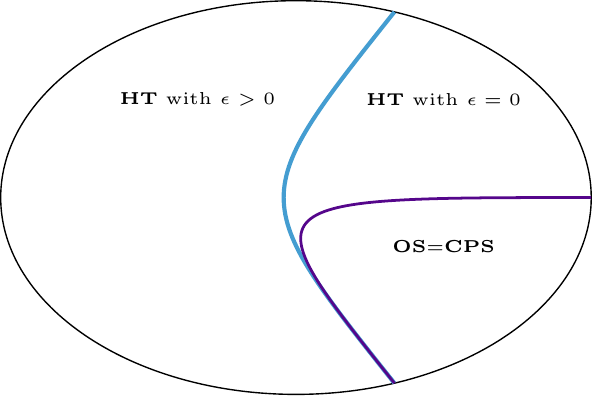}
      \caption{Relationship between OS, CPS, and HT.}\label{modelrelations}
         \end{figure}

The relationship between OS, CPS, and HT is illustrated in \autoref{modelrelations}.

\subsection{$\epsilon$-Ordered Surprises}

A natural way to generalize the Ordered Surprises representation is to retain the ``sequential selection'' of new beliefs while incorporating the idea of ``non-Bayesian reaction to unexpected events'' from the HT model. To do so, we introduce $\epsilon$-OS.  Since $OS$ is equivalent to CPS, we interpret $\epsilon$-OS as a way to the define a one-parameter non-Bayesian extension of the CPS. This extension may lead to an interesting, non-Bayesian generalization of sequential equilibria. 

\begin{defin} A complete updating rule $P$ has an $\epsilon$-\textbf{Ordered Surprise} representation if there are probability distributions $\mu_0, \ldots, \mu_K \in \Delta(S)$ and $\epsilon\in [0, 1)$ such that 
\[P(\cdot|E)=\text{BU}(\mu_{k^*}, E)\text{ where }k^*=\min\{k\le K \mid \mu_k(E)>\epsilon\},\] 
for every $E\in\Sigma$. 
\end{defin}

The $\epsilon$-OS representation incorporates the key idea of HT by allowing for non-Bayesian reactions to unexpected events:  $\mu_j(E)\leq\epsilon$. However, it provides additional structure to the posterior selection process. The $\epsilon$-OS remains a special case of HT.

\begin{theo}\label{eOS} Any complete updating rule that has an $\epsilon$-OS representation also has a HT representation. Moreover, if $\epsilon=0$, then the threshold for HT representation is also zero. 
\end{theo}

\section{Ordered Surprises Preferences}\label{sec:osp}

In this section, we provide a behavioral characterization of OS. Say that a family of preferences $\{\succsim_{E}\}_{E \in \Sigma }$ admits an \textbf{Ordered Surprises} (OS) representation if 
there is a complete updating rule $P$ that has an OS representation $\mathcal{O}=\langle \mu_0, \ldots, \mu_K \rangle$ and each preference $\succsim_{E}$ admits an SEU representation $(u_k, P(\cdot|E))$. That is, 
for all $f,g \in \cal F$:
\[f \succsim_{E_{}} g \quad \text{if and only if} \quad \sum_{s \in E_{}}  u_k\big(f(s)\big)P(s|E)  ~\geq~ \sum_{s \in E_{}}  u_k\big(g(s)\big) P(s|E).\]
 
Note that risk attitudes may depend on the order of the surprise, consistent with the approach taken by \citet{Myerson1986a}.

Our first axiom consists of several basic conditions that ensure an SEU representation. 

\begin{ax}[\bf{Conditional SEU Preferences}]\label{C-SEU} for every  $E \in \Sigma$, 
\begin{itemize}
\item[(i)] \textbf{\emph{(\text{Weak Order})}} $\succsim_E$ is complete and transitive; 
\item[(ii)] \textbf{\emph{(Continuity)}} for any $f, g, h\in F$, if $f\succ_E g$ and $g\succ_E h$, then there are $\alpha, \beta\in (0, 1)$ such that $\alpha f+(1-\alpha) h\succ_E g$ and $g\succ_E \beta f+(1-\beta) h$.

\item[(iii)] \textbf{\emph{(Independence)}} for any $f, g, h\in F$ and $\alpha\in (0, 1]$, 
\begin{equation*}
f\succsim_E g  ~\text{ if and only if }~ \alpha f+(1-\alpha) h\succsim_E \alpha g+(1-\alpha) h
\end{equation*}
\item[(v)]\textbf{\emph{(Nondegeneracy)}} there are $f, g\in \cal F$ such that $f\succ_E g$. 

\end{itemize}
\end{ax}  

Each condition is standard and so discussion is omitted. We also assume \nameref{Cons}, a well-known condition that ensures information is believed by the DM. 

\begin{ax}[\bf{Consequentialism}]\label{Cons} For every $E \in \Sigma$ and all $f, g \in \cal F$, if
 $f(s)=g(s)$  for all $s \in E$, then $f  \sim _{E} g$.
\end{ax}  

This axiom requires that each conditional preference $\succsim_{E}$ depends solely on the conditioning event $E$. Combined with \nameref{C-SEU},  \nameref{Cons} implies that all states in the complementary event, $S \setminus E$, are null according to $\succsim_{E}$. 

Our next axiom is novel and is the key condition for an OS representation. 

\begin{ax}[\bf{Conditional Consistency}]\label{CC} For all $E \in \Sigma$, $\succsim_{E}$-feasible $A \subset E$, and $f, g, h\in \cal F$,  \[f_A h \succsim_{E} g_A h ~\text{ if and only if }~ f\succsim_{A} g.\]
\end{ax}

\nameref{CC} implies Dynamic Consistency (DC) but also has bite on events that are (ex-ante) $\succsim_S$-null. For comparison, note that DC only has implications for events that are ex-ante feasible: for $A$ that is $\succsim_S$-feasible,  $f_A h \succsim_{S} g_A h ~\text{ if and only if }~ f\succsim_{A} g.$ In essence, \nameref{CC} extends the logic of DC to all conditional preferences $E$ and nested events that are $\succsim_E$-feasible (see \cite{ELeB93} and \cite{Gh2002} for an excellent discussion of DC).

\begin{theo}\label{OS-Foundation}
A family of preferences $\{\succsim_{E}\}_{E\in \Sigma}$ admits an OS representation if and only if it satisfies \nameref{C-SEU}, \nameref{Cons} and \nameref{CC}.
\end{theo}

\begin{ax}[\bf{Risk Independence}]\label{RI} For all lotteries $p, q\in \Delta(X)$, and events $E \in \Sigma$,
\begin{equation*}
p \succsim_S q \quad \text{if and only if} \quad p\succsim_E q.
\end{equation*}
\end{ax}

\begin{prop}\label{RI-prop}Suppose the family of preferences $\{\succsim_{E}\}_{E\in \Sigma}$ admits an OS representation. The family satisfies \nameref{RI} if and only if risk attitudes are surprise independent: for each $k>0$, there are $\alpha_k >0$, and $\beta_k$ such that $u_k=\alpha_k u_0 + \beta_k$. 
\end{prop}

\section{Lexicographic Probability System}\label{sec:lps}

When modeling conditional beliefs, the notion of sequential belief selection appears frequently in the literature. Accordingly, the word ``lexicographic" is often used to describe such procedures. However, ``lexicographic" is used with various meanings, as these models are typically quite different. Consequently, it is useful to clarify the difference between the Conditional LPS of \cite{BBD1991} and CPS, since we show that CPS precisely uses a sequence of probability distributions. 

\begin{defin}\label{LPS} A \textbf{Lexicographic Probability System} (LPS) is a list of probability distributions $(\mu_0, \ldots, \mu_K)$. The Conditional LPS (CLPS) given $E$ is the LPS obtained from the Bayesian updates of each $\mu_{k}$ (such that it is well-defined): $(BU(\mu_{k_1},E), \ldots, BU(\mu_{k_J},E))$. 
\end{defin}

The LPS Preference $\succsim^{LPS}$ is defined by \[ f \succsim^{LPS} g \quad \text{if and only if} \quad \left(\sum_{s\in S}  u\big(f(s)\big)\mu_i(s)\right)_{i=0}^{K} ~ \geq^L ~\left(\sum_{s \in S}  u\big(g(s)\big) \mu_{i}(s)\right)_{i=0}^{K}, \]
where $\geq^L$ is the lexicographic order on $\R^{K+1}$. The Conditional $\succsim_E$ is defined analogously. 

There are several differences between the CLPS and CPS.\footnote{\cite{HAMMOND1994} makes a mathematical connection between CPS and CLPS: some probability distributions generated by CPS also can be generated by CLPS. However, there is no behavioral connection between CLPS and CPS; there is no preference that has both CLPS and CPS representations (except SEU preferences with full support).} 
First, our axiomatic characterization (in addition to \cite{Myerson1986b}) shows that CPS is compatible with SEU, while CLPS preferences are not because they violate Continuity (except for $K=0$). Hence, from the behavioral point of view, CLPS and CPS are fundamentally different.\footnote{Imposing \nameref{RI} (State Independence in \cite{BBD1991}) on CLPS preferences implies full-support. In contrast, the CPS is a complete updating rule and allows for risk preferences to depend on the surprise order.} Second, since the CLPS is defined with Bayes' rule, it must rule out (Savage) null events to be well defined.\footnote{In an LPS with null events, the CLPS is not defined for these events. Thus, LPS updating is incomplete; there are scenarios in which some conditional probabilities are not well-defined. In contrast, the CPS is a complete updating rule; conditional probabilities are always well-defined.} In contrast, CPS allows for null events and beliefs are well-defined for all conditioning events.  

To further clarify, the following example illustrates that CLPS rules out indifference (indifferences are resolved ex-ante), while OS allows for information to resolve indifferences. That is, OS preferences exhibit a (strictly) positive value of information \citep{ELeB93,  Gh2002}: $f_Eh \sim g_Eh$ and $g \succ_E f$ for some $E$ and some acts.  
  
  \begin{exm}\label{ex: LPS} Since (Conditional) LPS violates continuity, it rules out indifference between distinct acts. Recall our introductory coin example and consider the following acts: 
\begin{align*}
f = \begin{cases} 
\$v & \text{if } s \in \{h,t\} \\ 
%\$w & \text{if } s \in C \\ 
\$0 & \text{otherwise}\end{cases} 
\, & \,
g =  \begin{cases} 
\$1 & \text{if } s \in \{h,t\} \\ 
\$t & \text{if } s \in \{e,e'\} \\ 
\$0 & \text{otherwise}\end{cases}. 
\end{align*}

Under OS, $f \sim^{OS} g$ if and only if $v=1$, $f \succ^{OS} g$ when $v>1$, and $g \succ^{OS} f$ if $v < 1$; the value of $t$ is irrelevant. In the LPS generated by $\mu_0,\mu_1,\mu_2$ the DM is never indifferent between $f$ and $g$. If $v>1$,  $f \succ^{LPS} g$ whereas if $v\leq 1$, $g \succ^{LPS} f$. 

Suppose $v=1$, and consider $E=\{e,e'\}$. Under OS, the realization of $E$ can resolve indifference: $f \sim^{OS} g$ and $g \succ^{OS}_E f$. In contrast, under CLPS the ex-ante and conditional preferences must agree: $g \succ^{LPS} f$ and $g \succ^{LPS}_E f$.  CLPS preferences can never exhibit resolution of indifference. 
 
\end{exm}

\section{Concluding Remarks}

We introduce and behaviorally characterize the Ordered Surprises representation of preferences. The OS belief structure yields a complete updating rule that is (behaviorally) equivalent to the CPS. Additionally, we show that OS is a special case of HT, clarifying the relationship between CPS and HT. 

The OS representation provides a tractable way to model any type of dynamic decision problem that allows for zero-probability contingencies, including dynamic games, strategic communication, information transmission, and Bayesian persuasion. Fruitful directions for future research include extending the idea of Ordered Surprises to ambiguity or other non-SEU preferences.

%Finally, we propose a one-parameter, non-Bayesian extension of CPS and show that it is still a special case of HT. 

\appendix

\section{Proofs}

\subsection{Proof of \autoref{OS-CPS}}

\noindent{($\Leftarrow$)} Consider an updating rule $P$ with an OS representation $\mathcal{O}=\langle \mu_0, \ldots, \mu_K\rangle$. We must show that $P$ satisfies the conditions in \autoref{cps}. By assumption, the rule is complete. 
 
 Since $P$ has an OS representation, let $\Sigma_0, \ldots, \Sigma_K$ be the partition of $\Sigma$ such that for each $k$, $\Sigma_k$ is the collection of events for which $\mu_k$ is used for updating: \[\Sigma_k=\{E\in \Sigma \mid k=\min\{\tilde{k}\le K \mid \mu_{\tilde{k}}(E)>0\}\}.\] 

We first show that $P$ is consequentialist. Pick any $E \in \Sigma$. Then for some $k^*$, $E \in \Sigma_{k^*}$. By assumption, $P(\cdot|E)=BU(\mu_{k^*}, E)$, which is well-defined. Hence $P(E|E)=1$.  

To show that \eqref{eq:cps} holds, take $G,F,E \in \Sigma$ such that $G \subset F \subset E$ and $F\neq \varnothing$. 

{\bf{Case 1:}} Suppose $G,F,E \in \Sigma_k$ for some $k \leq K$. Since $P$ has an OS representation, 
\begin{equation}
\begin{split}
P(G|E)=BU(\mu_k,E)(G)&=\frac{\mu_k(G)}{\mu_k(E)}=\frac{\mu_k(G)}{\mu_k(E)}\times\frac{\mu_k(F)}{\mu_k(F)} =  \\  \frac{\mu_k(G)}{\mu_k(F)} \times \frac{\mu_k(F)}{\mu_k(E)}
 = BU(\mu_k,F)(G) & \times BU(\mu_k,E)(F)= P(G|F)\times P(F|E).
\end{split}
\end{equation}

{\bf{Case 2}:} For $j \in \{G,F,E\}$, let $k_j$ denote the index for which $j \in \Sigma_{k_j}$. Note that $k_G=k_F=k_E=$ is precisely Case 1. If $P(G|E)>0$, then $k_G=k_E$. Further, since $G\subset F$, it follows that $0< \mu_{k_E}(G) \leq \mu_{k_E}(F)$, and so $P(F|E)>0$ and $k_G=k_F=k_E$.  Suppose $P(G|E)=0$. Then, $k_G \neq k_E$. For \eqref{eq:cps} to hold, we must show that either $P(G|F)$ or $P(F|E)$ is zero (or both). Suppose $P(F|E)>0$. Then, 
\[P(F|E)=BU(\mu_{k_E}, E)(F)=\frac{\mu_{k_E}(F)}{\mu_{k_E}(E)} >0.\]
Hence, $F \in \Sigma_{k_E}$, and so $k_E=k_F$. Consequently, $k_G \neq k_F$ and $P(G|F)=0$. Hence, \eqref{eq:cps} holds.

\noindent{($\Rightarrow$)}  Let $P$ be a CPS. We will construct an Ordered Surprises representation $\mathcal{O}=\langle \mu_0, \ldots, \mu_K \rangle$.

Set $S_0:=S$ and consider $P(\cdot|S_0) \in \Delta(S_0)$. Define $\Sigma_0 :=\{ E \subseteq S_0 \mid  P(E|S_0)>0\}$ and $F_0:=\{s \in S_0 : P(\{s\}| S_0)>0\}$. For each $E \in \Sigma_0$,  $P(\cdot |E)$ is the Bayesian update of $P(\cdot | S_0)$ conditional on $E$.  To see why, consider $E \in \Sigma_0$ and $G \subset S_0$. Since $G \cap E \subseteq E \subseteq S_0$, by \eqref{eq:cps}, we have $P(G \cap E | S_0) = P(G \cap E | E)\times P(E | S_0)$. Equivalently,
\begin{align}
\frac{P(G \cap E |  S_0)}{P(E| S_0)}&=P(G \cap E |  E),
\end{align}
since $P(E|S_0)>0$. Further, since $P$ is a CPS, $P(E | E)=1$. Hence, $P(\cdot|E)$ is the Bayesian update of $P(\cdot | S_0)$. Let $\mu_0:=P(\cdot|S_0)$. If $F_0=S_0$ (all states are non-null), there are no surprises and we are done. However, if $F_0 \subset S_0$, there are surprises and we proceed to the next step. 

Define $S_1:=S_0\setminus F_0$, the first-order surprise.  Consider $P(\cdot | S_1)$ and, as before, define the sets $\Sigma_1 :=\{ E \subseteq S_1 \mid  P(E|S_1)>0\}$ and $F_1:=\{s \in S_1 : P(\{s\}| S_1)>0\}$. Since $P$ is a CPS, $P(S_1 | S_1)=1$. By the same argument as above, for each $E \in  \Sigma_1$, $P(\cdot | E)$ is the Bayesian update of  $P(\cdot| S_1)$ given $E$. Let $\mu_1:=P(\cdot|S_1)$. If $F_1 = S_1$, there are no more surprises. If $F_1 \subset S_1$, we proceed.

In step $k$, we define $S_k=S_{k-1}\setminus F_{k-1}$. Consider the conditional probability $P(\cdot | S_k)$, and define $\Sigma_k =\{ E \subseteq S_k :  P(E| S_k)>0\}$ and $F_k=\{s\in S_{k} : P(\{s\} | S_{k})>0\}$. As before, $P(\cdot | E)$ must be the Bayesian update of $p(\cdot | S_k)$ conditional on $S_k$. We define $\mu_k:=P(\cdot|S_k)$. We proceed in this way until we can find $K \leq |S|$ such that $F_K=S_{K}$. Since $S$ is finite, such a $K$ exists.

Finally, notice that $\Sigma_0,\Sigma_1,\ldots \Sigma_K$ is a partition of $\Sigma$ and that $\mathcal{O}=\langle \mu_0, \ldots, \mu_K\rangle$ is an OS representation of $P$.

\subsection{Proof of \autoref{OS=HTE}}

Consider a complete updating rule $P$ that admits an OS representation $\mathcal{O}=\langle \mu_0, \ldots, \mu_K \rangle$. We will construct a second-order belief $\rho \in \Delta(\{\mu_0, \ldots, \mu_K\})$ such that $P$ all has a HT  representation. 

Without loss,  $\{supp(\mu_k)\}_{k=0}^K$ forms a partition of $S$.  For each $k\leq K$, let $F_k$ denote the feasible states for $\mu_k$ and $\delta_k =\min_{s \in F_k}\mu_{k}(s) \in (0,1)$. Define the following $K+1$ numbers: $v(0)=1$ and for $k\ge 1$, $v(k)=\frac{\delta_{k-1}}{2}v(k-1)$.  This is a decreasing sequence with $v(k)=2^{-k}\prod_{j=0}^{k-1}\delta_j $ for $k \ge 1$.  We use these numbers to construct $\rho$. For each $\mu_k$,  \[\rho(\mu_{S_k})= \frac{v(k)}{\sum_{j=0}^{K}v(j)},\] and $\rho(\mu)=0$ for all $\mu \in \Delta(S)\setminus \mathcal{O}$.  Finally, let $\epsilon=0$. 

By construction, $\{\mu_{0} \}= \argmax \rho(\mu)$. Consider any $E \in \Sigma$. If $\mu_0(E)>0$, then $P(\cdot|E)=BU(\mu_0, E)$ since $P$ also has an OS representation. If $E \subseteq F_k$ for some $k>0$, then since $\mu_{k}(E)>0$ and $\mu_{k'}(E)=0$ for all $k'\neq k$, it follows that $\mu_{k}(E)\rho(\mu_k) > 0 = \mu_{k'} \rho(\mu_{k'})$ and $P(\cdot|E)=BU(\mu_{k}, E)$. The final possibility is $E \cap F_k \neq \emptyset$ for multiple $k$.  Let $K(E)$ denote all $k$ such that $E \cap F_k \neq \emptyset$ and let $\underline{k}$ be the least such $k$. Note that $\mu_{\underline{k}}(E) \ge \delta_{\underline{k}}$ and if $k \in K(E) \setminus{\underline{k}}$, then $k > \underline{k}$ and 
 \[\frac{\rho(\mu_{\underline{k}})}{\rho(\mu_{k})}=\frac{v(\underline{k})}{v(k)} = \frac{2^{-\underline{k}}\prod_{j=0}^{\underline{k}-1}\delta_j }{2^{-k}\prod_{j=0}^{k-1}\delta_j }= \frac{2^{k-\underline{k}}}{\prod_{j=\underline{k}}^{k-1}\delta_j} .\]
Thus, 
\[ \prod_{j=\underline{k}}^{k-1}\delta_j \rho(\mu_{\underline{k}})   =2^{k-\underline{k}}\rho(\mu_{k}).\]
Since $\mu_{\underline{k}}(E) \ge \delta_{\underline{k}} \ge \prod_{j=\underline{k}}^{k-1}\delta_j$ and $\mu_{k}(E) \le 1$, it follows that  
\[ \mu_{\underline{k}}(E) \rho(\mu_{\underline{k}}) \geq \prod_{j=\underline{k}}^{k-1}\delta_j \rho(\mu_{\underline{k}}) = 2^{k-\underline{k}}\rho(\mu_{k}) > \mu_{k}(E) \rho(\mu_{k}),\]
and \[ P(\cdot|E)=BU(\mu_{\underline{k}}, E),\]
where the last line holds because $P$ admits an OS representation.

\subsection{Proof of \autoref{eOS}}

Take any complete updating rule $P$ that has an $\epsilon$-OS representation with some $\epsilon\in [0, 1)$. Then, there are probability distributions $\mu_0, \ldots, \mu_K$ such that 
\[P(\cdot|E)=\text{BU}(\mu_{k^*}, E)\text{ where }k^*=\min\{k\le K \mid \mu_k(E)>\epsilon\}\]
 for every $E\in\Sigma$. Let $\Sigma_0, \ldots, \Sigma_K$ be a partition of $\Sigma$ such that for each $k$, $\Sigma_k$ is the collection of events for which the prior $\mu_k$ is used for updating: \[\Sigma_k=\{E\in \Sigma \mid k=\min\{\tilde{k}\le K \mid \mu_{\tilde{k}}(E)>\epsilon\}\}.\] Throughout this proof, we assume that for any $k\le K$, $E_k$ is an element of $\Sigma_k$. Take $\overline{\rho}_0, \underline{\rho}_0, \ldots, \overline{\rho}_K, \underline{\rho}_K$ with \[\overline{\rho}_0>\underline{\rho}_0>\overline{\rho}_1>\underline{\rho}_1>\ldots>\overline{\rho}_K>\underline{\rho}_K>\delta\,\overline{\rho}_0>0\]
 and $\underline{\rho}_k>\overline{\rho}_k\,\mu^{E'_k}(E_k)$ for any $E_k, E'_k$ with $\mu^{E'_k}(E_k)<1$. 
 
 Let $\mu^E_k=\text{BU}(\mu_k, E)$ for any $E\in \Sigma$. Let $\rho$ be an element of $\Delta(\{\mu^{E_k}_k\}_{k\le K, E_k\in \Sigma_k})$ such that (i) $\rho(\mu ^{E_k}_k)\in (\underline{\rho}_k, \overline{\rho}_k)$ for any $k\le K$ and (ii) $\rho(\mu ^{E_k}_k)>\rho(\mu ^{E'_k}_k)$ if $\mu ^{E_k}_k\neq \mu ^{E'_k}_k$ and $\mu ^{E'_k}_k(E_k)=1$. 
 
 Let us first show that there is $\rho$ that satisfies (ii). Let $\mu ^{E_k}_k\succ^* \mu ^{E'_k}_k$ if $\mu ^{E_k}_k\neq \mu ^{E'_k}_k$ and $\mu ^{E'_k}_k(E_k)=1$. It is enough to show that $\succ^*$ is acyclic. To show acyclicity, suppose that there are $E^1_k, \ldots, E^T_k$ such that $\mu ^{E^t_k}_k(E^{t+1}_k)=1$ for each $t\le T-1$ and $\mu ^{E^T_k}_k(E^1_k)=1$. Note that  $\mu ^{E'_k}_k(E_k)=1$ is equivalent to $\text{supp}(\mu_k)\cap E'_k\subseteq E_k$. Hence, $\mu ^{E'_k}_k(E_k)=1$ implies $\text{supp}(\mu_k)\cap E'_k\subseteq \text{supp}(\mu_k)\cap E_k$. Then, $\mu ^{E^t_k}_k(E^{t+1}_k)=1$ implies $\text{supp}(\mu_k)\cap E^{t}_k\subseteq \text{supp}(\mu_k)\cap E^{t+1}_k$ and $\mu ^{E^T_k}_k(E^1_k)=1$ implies  $\text{supp}(\mu_k)\cap E^T_k\subseteq \text{supp}(\mu_k)\cap E^1_k$. Hence,  $\text{supp}(\mu_k)\cap E^t_k=\text{supp}(\mu_k)\cap E^{t'}_k$ for any $t, t'$; i.e., $\mu ^{E^t_k}_k=\mu ^{E^{t'}_k}_k$.    
 
 We now show that $P$ has a HT representation with $(\rho, \delta)$ when $\delta$ is large enough. Hence, we shall show that for any $E_k$, 
 
 \[\rho(\mu^{E_k}_k)\mu^{E_k}_k(E_k)=\rho(\mu^{E_k}_k)>\rho(\mu^{E_j}_j)\mu^{E_j}_j(E_k)\text{ for any }j\neq k.\]
For any $j>k$, the above holds since $\rho(\mu^{E_k}_k)>\rho(\mu^{E_j}_j)$. Suppose now $j<k$. In this case, $\mu_j(E_k)\le \epsilon$ since $k$ is the lowest index such that $\mu_k(E_k)>\epsilon$. Then, $\mu^{E_j}_j(E_k)=\text{BU}(\mu_j, E_j)(E_k)=\frac{\mu_j(E_k\cap E_j)}{\mu_j(E_j)}$. Since $\mu_j(E_k)\le \epsilon$ and $\mu_j(E_j)>\epsilon$, there is a large enough $\delta\in [0, 1)$ such that $\mu^{E_j}_j(E_k)\le \delta$. Hence, by the construction of $\rho$, 
 \[\rho(\mu^{E_k}_k)>\delta \rho(\mu^{E_j}_j)\ge \rho(\mu^{E_j}_j)\mu^{E_j}_j(E_k).\]
We finally show that the HT representation correctly chooses $\mu^{E_k}_k$ among $\{\mu^{E'_k}_k\}_{E'_k}$ for each $E_k$. When $\mu^{E'_k}_k(E_k)<1$, we have 
 \[\rho(\mu^{E_k}_k)\mu^{E_k}_k(E_k)=\rho(\mu^{E_k}_k)>\underline{\rho}_k>\overline{\rho}_k\mu^{E'_k}_j(E_k)>\rho(\mu^{E'_k}_k)\mu^{E'_k}_j(E_k).\]
When $\mu^{E'_k}_k(E_k)=1$ and $\mu^{E_k}\neq \mu^{E'_k}_k$, \[\rho(\mu^{E_k}_k)\mu^{E_k}_k(E_k)=\rho(\mu^{E_k}_k)>\rho(\mu^{E'_k}_k)=\rho(\mu^{E'_k}_k)\mu^{E'_k}_j(E_k).\]

It is immediate from the above construction of $\delta$, $\delta=0$ whenever $\epsilon=0$. 
%\end{proof}

\subsection{Proof of \autoref{OS-Foundation}}

Necessity of the axioms is trivial, so we only prove sufficiency. Take $E \in \Sigma$ and a conditional preference relation $\succsim_{E}$ . 

{\bf{Step 1.}} By \nameref{C-SEU}, $\succsim_{E}$ admits a state-dependent SEU representation. There are a probability measure $\mu_E$ on $S$ and, for each $s\in E$, an expected utility function $U_s$ on $\Delta(X)$, such that for all $f,g \in \cal F$:
\[f \succsim_{E} g \quad \text{if and only if} \quad \sum_{s\in E}  U_s\big(f(s)\big)\mu_E(s) ~ \geq~ \sum_{\omega \in S_E}  U_s\big(g(s)\big) \mu_E(s).\]
A state $s\in S$ is $\succsim_{E}$-null if and only if $\mu_E(s)=0$. By \nameref{Cons}, states outside of $E$ are $\succsim_{E}$-null. Moreover, $\mu_E$ is unique and for each $s \in E$, $U_s$ is unique up to a positive affine transformation. 

{\bf{Step 2.}} ($U_s$ is state-independent on $F_E$). Let $F_E=\{s \in S : s \text{ is }\text{$\succsim_{E}$-feasible}\}$ be the set of feasible states according to $\succsim_{E}$. Take $A=\{s\}$ where $s\in F_E$. By Step 1, $\succsim_{\{s\}}$ is SEU with respect to $U_s'$ and $\mu_{\{s\}}$ such that for all $f,g \in \cal F$, $f \succsim_{\{s\}} g$ if and only if $U_s'(f(s)) \geq U_s'(g(s))$ if and only if $U_s'(p) \geq U_s'(q)$ where $f(s)=p$ and $g(s)=q$.  Since $A=\{s\}$ is a $\succsim_{E}$-feasible state, by \nameref{CC}, for all lotteries $p,q \in \Delta(X)$, 
\[p_{\{s\}} h \succsim_{E} q_{\{s\}} h ~\text{ if and only if }~ p\succsim_{\{s\}} q,\]
equivalently,
\[U_s(p)\mu_E(s)  \succsim_{E} U_s(q)\mu_E(s)   ~\text{ if and only if }~ U_s'(p)  \succsim_{\{s\}} U_s'(q),\]
equivalently, by Step 1,
\[U_s(p) \geq U_s(q) ~\text{ if and only if }~ U_s'(p) \geq U_s'(q).\]
Hence, $U_s'$ is a positive affine transformation of $U_s$. Hence, for each $\succsim_{E}$-feasible state $s$ in  $F_E$, $U_s$ is state-independent and so without loss $U_s'=U_s$.

{\bf{Step 3.}} (Bayesian updating). Let $A \in \Sigma$ be a $\succsim_{E}$-feasible event with $A \subseteq E$. Notice that $A$ is $\succsim_{E}$-feasible if and only if $\mu_E(A)>0$, or equivalently, $A\cap F_E\neq \emptyset$. Consider acts $f_Ah$ and $g_Ah$ such that  $f_Ah \succsim_{E} g_Ah$. By Step 1,
\begin{eqnarray}
\sum_{s \in S}  U_s\big(f_Ah(s)\big)\mu_E(s) ~ \geq~ \sum_{s \in S}  U_s\big(g_Ah(s)\big) \mu_E(s).
\end{eqnarray}
By \nameref{Cons}, $\mu_E(S\setminus E)=0$. Thus,
\begin{eqnarray}
\sum_{s \in E}  U_s\big(f_Ah(s)\big)\mu_E(s) ~ \geq~ \sum_{s \in E}  U_s\big(g_Ah(s)\big) \mu_E(s),
\end{eqnarray}
or equivalently, 
\begin{eqnarray}\label{EQ-1-BU}
\sum_{s \in A}  U_s\big(f(s)\big)\mu_E(s) ~ \geq~ \sum_{s \in A}  U_s\big(g(s)\big) \mu_E(s).
\end{eqnarray}
By \nameref{CC}, $f\succsim_{A} g$. Thus, by Steps 1 and 2, 
\begin{eqnarray}\label{EQ-2-BU}
\sum_{s \in A}  U_s\big(f(s)\big)\mu_A(s) ~ \geq~ \sum_{s \in A}  U_s\big(g(s)\big) \mu_A(s).
\end{eqnarray}
Since Equations \eqref{EQ-1-BU} and \eqref{EQ-2-BU} are equivalent by \nameref{CC}, we have Bayesian updating
\begin{eqnarray}
\mu_A(s)=\frac{\mu_E(s)}{\mu_E(A)}~\text{for each}~s\in A.
\end{eqnarray}

{\bf{Step 4.}} (Ordered surprises). We proceed inductively. Let $S_0=S$, and consider $\succsim_{S_0}$. We let $\mu_0=\mu_{S_0}$, and if $\mu_0$ has full support, stop. Otherwise, let $S_1$ denote the set of $\succsim_{S_0}$-null states, and consider $\succsim_{S_1}$. Then, let $\mu_1=\mu_{S_1}$. By \nameref{Cons}, $\mu_1(S_0\setminus S_1)=0$. If $supp(\mu_1)=S_1$, stop. Otherwise, let $S_2$ denote the set of $\succsim_{S_1}$-null states. We proceed in this fashion until we reach a $K$ such that $supp(\mu_K)=S_K$. Since $S$ is finite, we must eventually stop. 

Then $\mathcal{O}=\langle \mu_0, \ldots, \mu_K \rangle$ constitutes an OS representation. To see why, note that by construction $supp(\mu_k)\cap supp(\mu_{k'})=\varnothing$ if $k\neq k'$ and $\cup_{k=0}^K supp(\mu_k)=S$. Further, from the preceding step, it follows that for every $E$, \[P(\cdot|E)=\mu_E =BU(\mu_{k^*}, E)= BU(\mu_{S_{k^*}}, E)\] where $k^*=\min\{k\le K \mid \mu_k(E) > 0\}$.

{\bf{Step 5.}} Combining steps, it follows that for each $k$ there is a $u_k$ such that $\succsim_E$ is represented by $(u_{k^*},P(\cdot|E))$, where $P(\cdot|E)=BU(\mu_k^*,E)$ and $k^*=\min\{k\le K \mid \mu_k(E) > 0\}$.

\subsection{Proof of \autoref{RI-prop}}

Given an OS representation, fix $k>0$ and consider some $E$ with $\mu_k(E) > 0$. By \nameref{RI}, it follows that for every $p,q \in \Delta(X)$, $p\succsim q$ iff $p \succsim_E q$. Consequently, $u_0(p) \geq u_0(q)$ iff $u_k(p) \geq u_k(q)$, and so by standard results $u_0$ and $u_k$ represent the same risk preferences.

\bibliographystyle{ecta}
\bibliography{OS_References}

\end{document}